\documentclass{ifacconf}

\usepackage{cite}
\usepackage{amsmath, amssymb, bm}
\usepackage{booktabs} 
\usepackage{color}

\usepackage{graphicx}      
\usepackage{natbib}  

\begin{document}
\begin{frontmatter}

\title{Quantized Output Feedback Stabilization 
	by Luenberger Observers\thanksref{footnoteinfo}} 

\thanks[footnoteinfo]{M. Wakaiki acknowledges The Telecommunications
	Advancement Foundation for their support of this work.}

\author{Masashi Wakaiki, }
\author{Tadanao Zanma, and}
\author{Kang-Zhi Liu}

\address{ Department of Electrical and Electronic 
	Engineering, Chiba University,
	Chiba 263-8522, Japan (email:{\tt  \ \{wakaiki,zamma\}@chiba-u.jp}, 
	{\tt kzliu@faculty.chiba-u.jp}).}

\begin{abstract}                
We study a stabilization problem for systems with
quantized output feedback.
The state estimate from a Luenberger observer is used 
for control inputs and quantization centers.
First we consider the case when only the output is quantized and
provide data-rate conditions for stabilization.
We next generalize the results to the case where
both of the plant input and output are quantized and where controllers
send the quantized estimate of the plant output to encoders as quantization centers.
Finally, we present the numerical comparison of
the derived data-rate conditions with those in the earlier studies
and a time response of an inverted pendulum. 
\end{abstract}


\end{frontmatter}

\section{Introduction} 
Control loops in a practical network contain channels
over which only a finite number of bits can be transmitted.
Due to such limited transmission capacity, 
we should quantize data before sending them out through a network.
However, large quantization errors lead to the deterioration of control performance.
One way to reduce quantization errors under data-rate constraints is
to exploit output estimates as quantization centers.
In this paper, we adopt Luenberger observers as output estimators due 
to their simple structure and aim to 
design an encoding strategy for stabilization.

A fundamental limitation of data rate for stabilization was first obtained by
\cite{Wong1999}, and
inspired by this result, data-rate limitations were studied for
linear time-invariant systems in \cite{Tatikonda2004},
for stochastic systems in \cite{Nair2004}, and
for uncertain systems in \cite{Okano2014}.
Although the so-called zooming-in and zooming-out encoding method
developed in \cite{Brockett2000, Liberzon2003} provides
only sufficient conditions for stabilization, this encoding procedure
is simple and hence was extended, e.g., 
to nonlinear systems in \cite{Liberzon2005, Liberzon2006},
to systems with external disturbances in \cite{Liberzon2007,Sharon2008,Sharon2012}, and
recently to switched/hybrid systems in \cite{Liberzon2014,Yang2015ACC,Wakaiki2016TAC}.
Readers are referred to the survey papers by \cite{Nair2007, Ishii2012},
and the books by \cite{Matveev2009,Liberzon2003Book}
on this topic for further information.

Although Luenberger observers has been widely used for
quantized output feedback stabilization, e.g., in
\cite{Liberzon2003Automatica, Ferrante2014ECC, Xia2010},
state estimates was exploited only to generate 
control inputs, and a quantization center was the origin.
However, to reduce quantization errors, output estimates play an 
important role as quantization centers.

The notable exception is the studies by \cite{Sharon2008,Sharon2012}.
The class of observers in
these studies covers
a Luenberger observer whose estimate is initialized by a pseudo-inverse observer, and
\cite{Sharon2008,Sharon2012}
provided a sufficient condition for stabilization with
unbounded disturbances. However,
this condition is not easily verifiable for the case of Luenberger observers.
Furthermore, these studies placed
assumptions that
input quantization is ignored and that
encoders contain state estimators for
sharing quantization centers with controllers.

%

In this paper, we present an output encoding method for the stabilization of
sampled-data systems with
discrete-time Luenberger observers.
The proposed encoding method is based on the
zooming-in technique
and employs estimates generated from a Luenberger observer
for both stabilization and quantization.
First we consider 
only output quantization and assume that
encoders also contain estimators.
Simple sufficient conditions for stabilization are obtained in the both case
of general Luenberger observers and of deadbeat observers.

Second we generalize the results of general Luenberger observers
to the situation where both of the plant input and output are quantized.
Moreover, in the second case, 
encoders do not estimate the plant state by themselves, but
controllers send the quantized estimate of the plant output 
to the encoders.
In contrast with the quantization of the plant output,
we quantize the plant input and the output estimate by using the origin
as the quantization center, which reduces computational resources
in the components of the plant side.
We see that if the closed-loop system without quantization is stable, then
there exists an encoding method such that the closed-loop system in the presence 
of three types of quantization errors
is also stable.

This paper is organized as follows.
In the next section, 
we study the case when only the plant output is quantized
and obtain two data-rate conditions for general Luenberger observers
and deadbeat observers.
In Section 3,
the proposed encoding method is extended to 
the case when both of the plant input and output are quantized.
Section~4 is devoted to the numerical comparison of the obtained data-rate conditions
with those of the earlier studies by \cite{Liberzon2003,Sharon2008,Sharon2012}
and the time response of an inverted pendulum.
We provide concluding remarks in Section 5.

\subsubsection*{Notation and Definitions:}
The symbol $\mathbb{Z}_+$ denotes the set of nonnegative integers.
Let $\lambda_{\min}(P)$ and $r(P)$ denote 
the smallest eigenvalue and the spectral radius
of $P \in \mathbb{R}^{\sf n\times n}$, respectively.
For a vector $v = [v_1~\!\dotsb ~\!v_{\sf n} ]^{\top} \in \mathbb{R}^{\sf n}$,
we denote 
its maximum norm by $|v| = \max\{|v_1|,\dots, |v_{\sf n}|\}$ and
the corresponding induced norm of a matrix $M \in \mathbb{R}^{\sf m\times n}$ by
\[\|M\|  = \sup \{  |Mv |_{\infty}:~
v\in \mathbb{R}^{\sf n},~|v|_{\infty}= 1 \}.
\]

\subsubsection*{Plant:}
We consider a continuous-time linear system
\begin{equation}
	\label{eq:plant}
	\Sigma_P: \begin{cases}
		\begin{aligned}
			\dot x(t) &= Ax(t) + Bu(t) \\
			y(t) &= Cx(t),
		\end{aligned}
	\end{cases}
\end{equation}
where $x(t) \in \mathbb{R}^{\sf{n}}$ is the state,
$u(t) \in \mathbb{R}^{\sf{m}}$ is the control input, and
$y(t) \in \mathbb{R}^{\sf{p}}$ is the output. 
This plant is connected with a controller through 
a time-driven encoder and zero-order hold (ZOH) with period $h >0$.
Define 
\[
x_k := x(kh),\qquad y_k := y(kh)
\]
for every $k \in \mathbb{Z}_+$, and also set
\begin{equation}
	\label{eq:discretization}
	A_d := e^{Ah},\qquad B_d := \int^{h}_0 e^{As}Bds.
\end{equation}

Throughout this paper, we place the following assumptions:
\begin{assum}[Initial state bound]
	\label{assum:initial_bound}
	A constant $E_{st} > 0$ satisfying $|x(0)| \leq E_{st}$ is known.
\end{assum}
\begin{assum}[Stabilizability and detectability]
	\label{assump:stabilizability_detectability}
	The discretized system $(A_d,B_d,C)$ is stabilizable and detectable.
\end{assum}
\begin{rem}
	We can obtain an initial state bound $E_{st}$ by the ``zooming-out'' procedure
	in \cite{Liberzon2003}.
\end{rem}

\section{Output Quantization}
In this section, we consider the scenario where
only the output $y_k$ is quantized and where
the encoder has computational resources to estimate the plant state.

\subsection{Controller}
\begin{figure}[tb]
	\centering
	\includegraphics[width = 7.5cm,clip]{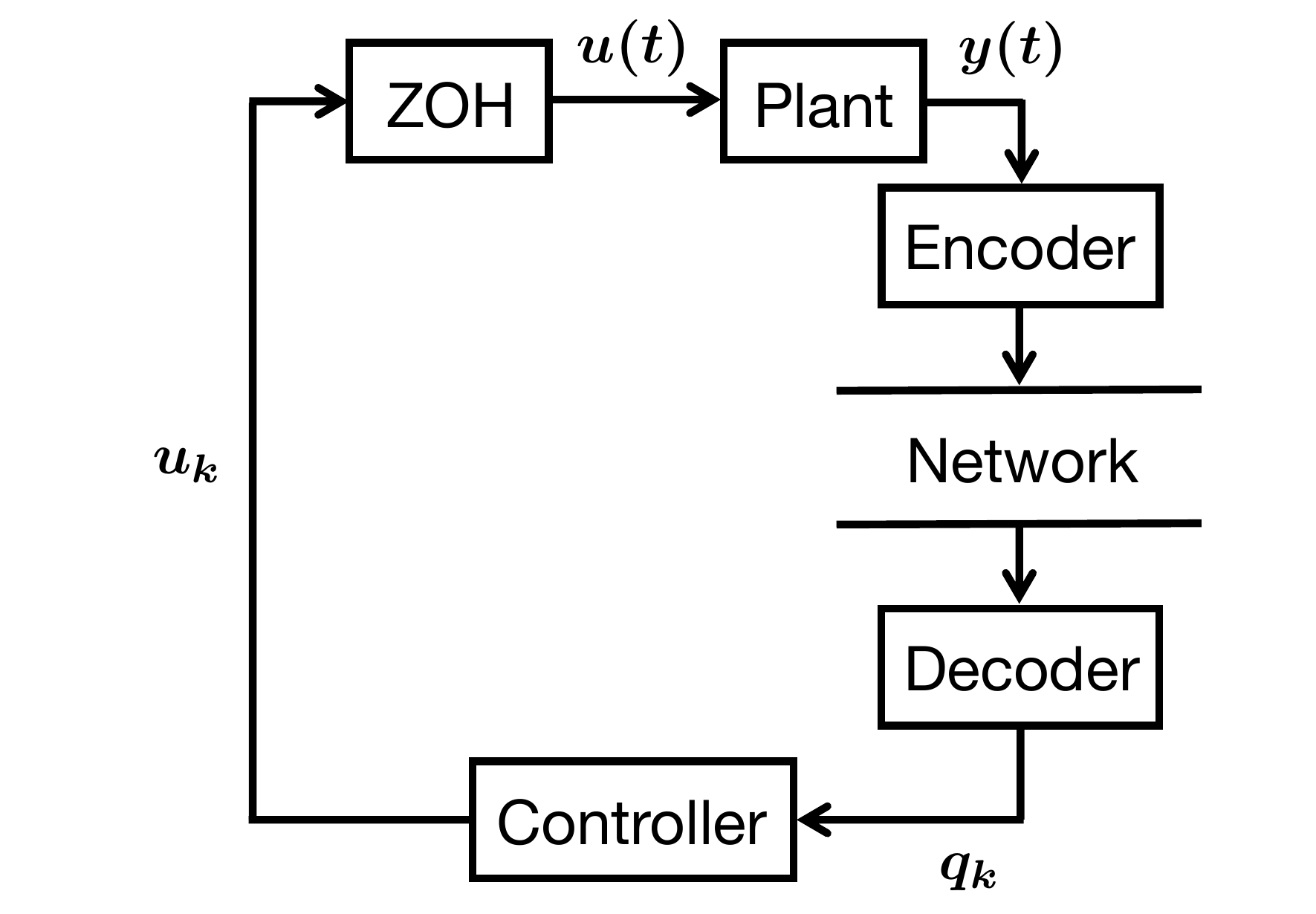}
	\caption{Closed-loop system with output quantization.}
	\label{fig:closed_loop}
\end{figure}

Let
$q_k \in \mathbb{R}^{\sf{p}} $ be the quantized value of the sampled output $y_k$ and
$K \in \mathbb{R}^{\sf{n} \times \sf{m}},
L \in \mathbb{R}^{\sf{n} \times \sf{p}}$ be a feedback gain and 
an observer gain, respectively.
For the plant $\Sigma_P$ in \eqref{eq:plant},
we use a discrete-time Luenberger observer for feedback control
and output quantization:
\begin{equation}
	\label{eq:observer}
	\Sigma_C: \begin{cases}
		\begin{aligned}
			\hat x_{k+1} &= A_d\hat x_k + B_d u_k + L(q_k - \hat y_k)  \\
			\hat y_k &= C \hat x_k \\
			u_k &= -K \hat x_k
		\end{aligned}
	\end{cases}
\end{equation}
where $\hat x_k \in \mathbb{R}^{\sf{n}}$ is the state estimate and 
$\hat y_k \in \mathbb{R}^{\sf{p}} $ is the output estimate.
We set the initial estimate $\hat x_{0}$ to be $\hat x_{0} = 0$. 
Each of the encoder and the controller contains the above Luenberger observer, and 
those observers are assumed to be synchronized.
Through the zero-order hold,
the control input $u(t)$ is generated as
\begin{equation*}
	u(t) = u_k \qquad (kh \leq t < (k+1)h).
\end{equation*}
Fig.~\ref{fig:closed_loop} shows the closed-loop system with quantized output.

\subsection{Output Encoding}
Suppose that we obtain an error bound $E_{k}$ such that $|y_k - \hat y_k| \leq E_{k}$.
The next subsection is devoted to the computation of a bound sequence $\{E_k\}$
for stabilization.

For each $k \in \mathbb{Z}_+$,
we divide the hypercube 
\begin{equation}
	\label{eq:quantization}
	\left\{ y \in \mathbb{R}^{\sf p}:~| y  - \hat y_k|_{\infty} 
	\leq E_{k} \right\}
\end{equation}
into $N^{\sf p}$ equal boxes and assign a number in $\{1,\dots, N^{\sf p}\}$
to each divided box by a certain one-to-one mapping.
Since $\hat x_0 = 0$,
we see from Assumption~\ref{assum:initial_bound}
that the error $e_k := x_k - \hat x_k$ satisfies
$|e_{0}|  = |x(0)| \leq E_{st}$.
Thus we can set \[
E_{0} := \|C\| E_{st}.
\]

The encoder sends to the controller
the number $\bar q_{k}$ of the divided box containing 
$y_k$, and then
the controller generates $q_{k}$ equal to the center of the box
with number $\bar q_{k}$.
If $y_k$ lies on the boundary on several boxes, then
we can choose any one of them.
This encoding strategy leads to
\begin{equation}
	\label{eq:qe_y}
	|y_k - q_k| \leq \frac{E_k}{N} =: \mu_k.
\end{equation}


\subsection{Computation of Bound Sequence $\{E_k\}$}
Here we obtain bound sequences $\{E_k\}$ 
and data-rate conditions for stabilization.
We first consider general Luenberger observers and next
focus on deadbeat observers.
\subsubsection{Use of General Luenberger Observers:}
The proposed encoding strategy with the following bound sequence $\{E_k\}$
achieves
the exponential convergence of the state under a certain data-rate condition.
\begin{thm}
	\label{thm:no_disturbances_non_deadbeat}
	Let Assumptions \ref{assum:initial_bound} and 
	\ref{assump:stabilizability_detectability} hold.
	Define the matrices $R$ and $\bar R$ by 
	\begin{equation}
		\label{eq:R_def}
		R:=A_d-LC,\qquad \bar R:= A_d-B_dK.
	\end{equation}
	Let
	the observer gain $L$ and the feedback gain $K$
	satisfy $r(R) < 1$, $r(\bar R) < 1$, and
	\begin{equation}
		\label{eq:CRL_bound}
		\left\|CR^{\ell}\right\| \leq M_0 \rho^{\ell},\quad 
		\left\|CR^{\ell}L\right\| \leq M \rho^{\ell}.
	\end{equation}
	for some $M_0,M> 0$ and $\rho < 1$.
	If we pick $N \geq 2$ so that 
	\begin{equation}
		\label{eq:bit_rate_cond_non_deadbeat_non_dis}
		\frac{M}{1-\rho}　 < N,
	\end{equation}
	then the proposed encoding method with
	a bound sequence $\{E_k\}$ defined by
	\begin{equation}
		E_{k+1} :=
		\begin{cases}
			M_0 E_{st} \rho  + \frac{M}{N}E_0& k = 0 \\
			\left(
			\rho + 
			\frac{M}{N}
			\right)
			E_k & k \geq 1.
		\end{cases}
		\label{eq:E_def_kal_qy}
	\end{equation}
	achieves the exponential convergence of 
	the state $x$ and the estimate $\hat x$.
\end{thm}

\begin{pf}
	The proof consists of two steps:
	\begin{itemize}
		\item[1)]
		Obtain the error bound $E_{k+1}$ from $E_0,\dots,E_k$.
		\item[2)]
		Show state convergence.
	\end{itemize}
	We break the proof of 
	Theorem \ref{thm:no_disturbances_non_deadbeat} into
	the above two steps.

	1)
	First we obtain an error bound
	$E_{k+1}$ for every $k\geq 0$ under
	the assumption that 
	$\mu_{0},\dots,\mu_k$ are obtained.
	
	Since the estimation error $e_k = x_k - \hat x_k$ satisfies
	\begin{equation}
		\label{eq:e_dynamics}
		e_{k+1} = 
		Re_k + L(y_k - q_k),
	\end{equation}
	and hence
	\begin{align}
		e_{k+1} 
		=
		R^{k+1} e_{0} +
		\sum^{k}_{\ell=0}R^\ell L (y_{k-\ell}-q_{k-\ell}).
		\label{eq:e_dynamics_IS}
	\end{align}
	Define $E_{k+1}$ by
	\begin{equation}
		\label{eq:E_k_def_gy}
		E_{k+1} := M_0 E_{st} 
		\rho^{k+1} 
		+
		M
		\sum_{\ell=0}^k \rho^{\ell} \mu_{k-\ell}
	\end{equation}
	for every $k \geq 0$.
	Then we conclude from \eqref{eq:e_dynamics_IS} that 
	\begin{equation}
		\label{eq:y_hy}
		|y_{k+1} - \hat y_{k+1}| \leq E_{k+1}.
	\end{equation} 
	Moreover,
	from \eqref{eq:E_k_def_gy}, we see that
	\begin{equation*}
		E_{k+1} - \rho E_k =
		\frac{M}{N}E_k
	\end{equation*}
	for every $k \geq 1$,
	and hence \eqref{eq:E_def_kal_qy} is obtained.
	Thus if \eqref{eq:bit_rate_cond_non_deadbeat_non_dis} holds,
	then $E_k$ and $\mu_k = E_k/N$ 
	exponentially converge to zero.
	\hspace*{\fill} $\blacksquare$
\end{pf}
and hence
\begin{equation}
	\label{eq:y_q_diff}
	| y_{k+1} - q_{k+1} | \leq \frac{E_{k+1}}{N}.
\end{equation}
By definition, $\mu_k = E_k/N$ satisfies
\begin{equation*}
	\mu_{k+1} - \rho \mu_k =
	\frac{M}{N} \mu_{k},
\end{equation*}
and hence
\begin{equation*}
	\mu_{k+1} =
	\left(\frac{M}{N} + \rho \right) \mu_{k}.
\end{equation*}
Thus if \eqref{eq:bit_rate_cond_non_deadbeat_non_dis} holds,
then $\mu_k$ exponentially converges to zero.

2)
Using the convergence of $\mu$, we next show the state convergence.
For every $k \geq 0$,
$\mu_k$ satisfies $\mu_k \leq \widetilde \mu_{0} \widetilde \rho^{k} $,
where $\widetilde \rho := M/N + \rho < 1$ and $\widetilde \mu_0 := 
\max\{\mu_0,\mu_1/\widetilde \rho\}$.
Then,
from \eqref{eq:qe_y} and \eqref{eq:e_dynamics_IS},
we have some constant $M_{e} > 0$ satisfying
$|e_k| < M_{e} \widetilde \mu_{0} \widetilde \rho^{k}$ for all $k \geq 0$.
Here we used $\|C\| \cdot |e_0| \leq E_{0} = N \mu_{0} \leq N\widetilde \mu_{0}$.

Since $\bar R = A_d - B_dK$ is Schur stable, 
there exist a positive scalar $c$ and
a positive definite matrix $P$ such that
\begin{equation*}
	\bar R^{\top} P \bar R - P \leq -cP.
\end{equation*}
Since
\begin{align}
	x_{k+1} &= A_dx_k - B_dK \hat x_k 
	= \bar R x_k + B_dKe_k, \label{eq:state_dynamics_error}
\end{align}
it follows that
\begin{align*}
	V(x_{k+1}) \!-\! V(x_k) 
	& \!\leq\! -cV(x_k) \!+\! 2 \|K^{\top}B_d^{\top}P\bar R\|_2 
	\! \cdot \! |x_k|_2 \!\cdot\! |e_k|_2 \\
	&\qquad+ \!
	\|K^{\top}B_d^{\top}P B_dK\|_2 \!\cdot\! 
	|e_k|_2^2.
\end{align*}
Young's inequality leads to
\begin{align*}
	2  |x_k|_2 \cdot |e_k|_2 \leq \frac{1}{\theta}|x_k|_2^2 + \theta|e_k|_2^2
\end{align*}
for all $\theta > 0$, and hence
\begin{equation}
	\label{eq:Lyap_ineq}
	V(x_{k+1}) \leq 
	\omega
	V(x_k) + \bar M_e |e_k|_2^2,
\end{equation}
where
\begin{align}
	\omega &:=
	1-
	c + \frac{\|K^{\top}B_d^{\top}P\bar R\|_2 }{\theta \lambda_{\min}(P)} \label{eq:omega_def}\\
	\bar M_e &:= \theta\|K^{\top}B_d^{\top}P\bar R\|_2 + \|K^{\top}B_d^{\top}P B_dK\|_2.
	\notag 
\end{align}
We choose a sufficiently large $\theta > 0$ so that 
$\omega < 1$.

Since \eqref{eq:Lyap_ineq} leads to
\begin{equation*}
	V(x_{k+1}) \leq 
	\omega^{k+1}
	V(x_0) +
	\bar M_e
	\sum_{\ell=0}^k
	\omega^{k-\ell}
	|e_{\ell}|^2_2
\end{equation*}
and since
$|e_k|_2 \leq \sqrt{\sf n} |e_k| \leq \sqrt{\sf n} M_e  \widetilde 
\mu_{0} \widetilde \rho^{k}$, we obtain
\begin{align*}
	V(x_{k+1}) \leq \omega^{k+1} 
	V(x_{0})   + 
	{\sf n}\bar M_e (M_e  \widetilde \mu_{0})^2 
	\sum_{\ell=0}^{k} \omega^{k-\ell}  \widetilde  \rho^{2\ell}.
\end{align*}
If $\omega \not= 2 \widetilde \rho$, then
\begin{align*}
	\sum_{\ell=0}^{k} \omega^{k-\ell}  \widetilde \rho^{2\ell}
	\leq
	\frac{\omega^{k+1} -  \widetilde \rho^{2(k+1)}}{\omega- \widetilde  \rho^2}
\end{align*}
otherwise,
\begin{align*}
	\sum_{\ell=0}^{k} \omega^{k-\ell}  \widetilde \rho^{2\ell}
	\leq
	(k+1)\omega^{k}.
\end{align*}
For every $\widetilde \omega$ with $\widetilde \omega > \omega$,
there exists a constant $M_{\tilde \omega} > 0$ such that
$M_{\tilde \omega} \widetilde \omega^k > k \omega^k$.
We therefore have
\begin{equation*}
	|x_k| \leq M_x  \widetilde \mu_{0} \gamma^{k}
\end{equation*}
for some $M_x > 0$, where $\gamma := \max \{ \widetilde \omega/2,  
\widetilde \rho\}$.
Here we again used $\|C\| \cdot |x_{0}| \leq N \widetilde \mu_{0} $.
Hence $\hat x_k$ satisfies
\begin{equation*}
	| \hat x_k | 
	\leq 
	| x_k | + | e_k|
	\leq M_{\hat x}  \widetilde \mu_{0} \gamma^{k} 
\end{equation*}
for some $M_{\hat x} > 0$.

Finally, $x$ satisfies
\begin{equation*}
	\dot x(t) = Ax(t) - BK \hat x_{k}
\end{equation*}
for all $t \in [kh, (k+1)h)$. From this linearity,
there exists $M >0$ such that
\begin{equation*}
	|x(t)| \leq M  \widetilde \mu_{0} e^{-\sigma t},
\end{equation*}
where
$\sigma := 1/(2h) \cdot \log(1/\gamma)$.
This completes the proof.



\subsubsection{Use of Deadbeat Observers:}
In the rest of
this section, we focus on
deadbeat observers. 
If the pair $(C, A_d)$ is observable,
then there exists a matrix 
$L \in \mathbb{R}^{{\sf n} \times {\sf p}}$ such that 
\begin{equation}
	\label{eq:R_property}
	R^{\eta} = (A_d - LC)^\eta = 0,
\end{equation}
where $\eta$ is the observability index of $(C, A_d)$.
Construction methods of such an observer gain $L$ have been developed for
deadbeat control; see, e.g., Chapter.~5 of \cite{Oreilly1983}.
Using the property \eqref{eq:R_property}, 
we obtain an alternative error bound sequence $\{E_k\}$ for stabilization.
\begin{thm}
	\label{thm:no_disturbances_qy}
	Let Assumptions \ref{assum:initial_bound} and 
	\ref{assump:stabilizability_detectability} hold.
	Assume that $(C,A_d)$ is observable.
	Define the matrices $R$ and $\bar R$ as in \eqref{eq:R_def}, and
	let
	the observer gain $L$ and the feedback gain $K$
	satisfy $R^{\eta} = 0$ and $r(\bar R) < 1$, where
	$\eta$ is the observability index of $(C,A_d)$.
	Set a constant $\alpha_{\ell}$ to be
	\begin{equation}
		\label{eq:alpha_def}
		\alpha_{\ell} := \frac{\left \|C R^{\ell} L \right\|}{N}
	\end{equation}
	for $\ell =0,\dots, \eta-1$.
	If we pick $N \geq 2$ so that a matrix $F$ defined by
	\begin{equation}
		\label{eq:F_def_deadbeat_y}
		F := 
		\begin{bmatrix}
			\alpha_0 & \alpha_1 & \dots & \alpha_{\eta-1} \\
			1 &  &  & 0 \\
			& \ddots &  &  \\
			0 & & 1 & 0
		\end{bmatrix}
	\end{equation}
	satisfies
	\begin{equation}
		\label{eq:F_spectrum}
		r(F) < 1,
	\end{equation}
	then the proposed encoding method with a bound sequence $\{E_k\}$ 
	defined by
	\begin{equation}
		\label{eq:E_def_deadbeat}
		E_{k+1} \!:=\!
		\begin{cases}\!
			\left\|CR^{k+1}\right\|E_{st} +
			\sum_{\ell = 0}^{k} \frac{\left\|CR^{\ell} L \right\|}{N} E_{k-\ell}
			& 0\!\leq\! k \!\leq\! \eta\!-\!2 \\
			\sum_{\ell = 0}^{\eta-1} \frac{\left\|CR^{\ell} L\right \| }{N}E_{k-\ell}
			& k \!\geq\! \eta \!-\!1
		\end{cases}
	\end{equation}
	achieves the exponential convergence of
	the state $x$ and the estimate $\hat x$.
\end{thm}

\begin{pf}
	Since $R$ satisfies $R^{\eta} = 0$, it follows from
	\eqref{eq:e_dynamics_IS} that
	for all $k \geq 0$, $E_{k+1}$ defined as in \eqref{eq:E_def_deadbeat}
	satisfies \eqref{eq:y_hy}.
	Note that $E_{k+1}$ with $k \geq \eta - 1$
	can be determined only from $\mu_{k-\eta+1},\dots,\mu_{k}$.
	
	Define a vector ${\bm \mu}_k$ by
	\begin{equation*}
		{\bm \mu}_k := 
		\begin{bmatrix}
			\mu_k \\ \vdots \\ \mu_{k-\eta+1}
		\end{bmatrix}
	\end{equation*}
	and a matrix $F$ by \eqref{eq:F_def_deadbeat_y}.
	Then it follows from \eqref{eq:E_def_deadbeat} that
	\begin{equation}
		\label{eq:bmu_eq}
		{\bm \mu}_{k+1} = F {\bm \mu}_k
	\end{equation}
	for all $k \geq \eta - 1$.
	Thus $\mu_k$ exponentially decreases to zero 
	if and only if $F$ is Schur stable.
	The rest of the proof is the same as that of 
	Theorem \ref{thm:no_disturbances_non_deadbeat},
	and hence we omit it.
	\hspace*{\fill} $\blacksquare$
\end{pf}

\begin{rem}
	As \eqref{eq:E_def_kal_qy},
	\eqref{eq:E_def_deadbeat} has also 
	the form of linear time-invariant recursion.
\end{rem}

\begin{rem}
	\label{rem:psuedo_inverse_ob}
	\cite{Sharon2008,Sharon2012} proposed
	the quantizer based on a psuedo-inverse observer.
	That quantizer
	achieves
	the closed-loop stability if
	$N \geq 2$ satisfies
	\begin{equation}
		\label{eq:output_liberzon}
		\left\|CA_d{\bf C}^{\dagger}\right\|< N,
	\end{equation}
	where 
	\begin{equation*}
		{\bf C} := 
		\begin{bmatrix}
			C \\ CA_d \\ \vdots \\ CA_d^{\eta-1}
		\end{bmatrix}
		A_d^{-\eta+1},\quad
		{\bf C}^{\dagger} :=
		\left(
		{\bf C}^{\top} {\bf C}
		\right)^{-1}
		{\bf C}^{\top},
	\end{equation*}
	and the total data size is $N^{\sf p}$.
	In the state feedback case of \cite{Liberzon2003}, the counterpart 
	of \eqref{eq:output_liberzon} is
	\begin{equation}
		\label{eq:state_liberzon}
		\left\|A_d \right\| < N,
	\end{equation}
	and the total data size is $N^{\sf n}$.
\end{rem}

\section{Input and Output Quantization}
In this section, we quantize both of the plant input and output.
Moreover, 
we assume in the previous section that
the encoder has computational resources to estimate the plant state, 
while we here study the scenario where
the controller sends the quantized output estimate to the encoder. 
Hence the encoder does not need to compute or store the estimate.

\subsection{Controller}
\begin{figure}[tb]
	\centering
	\includegraphics[width = 7.5cm,clip]{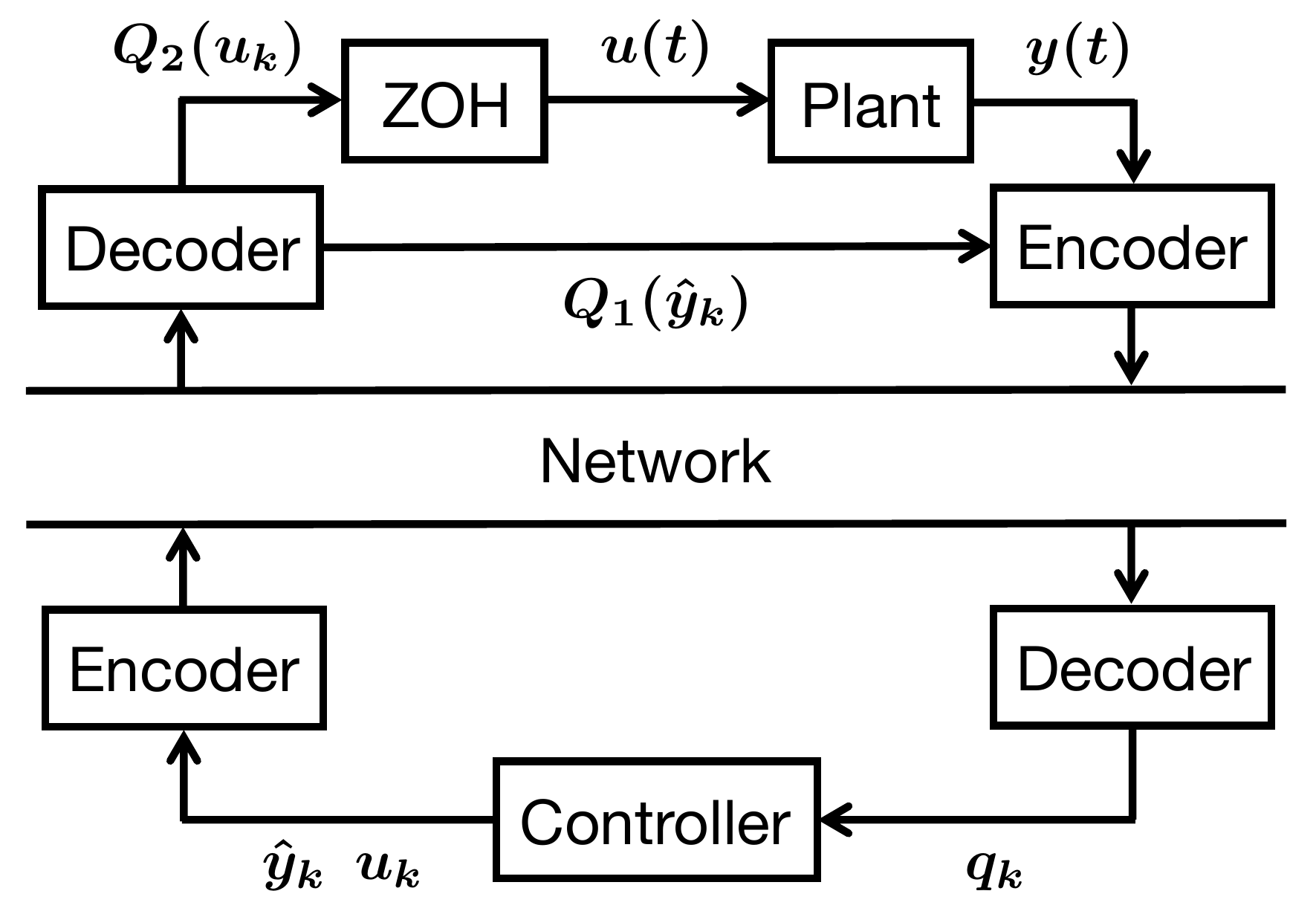}
	\caption{Closed-loop system with input and output quantization.}
	\label{fig:closed_loop_AQ}
\end{figure}

Let
$K \in \mathbb{R}^{\sf{n} \times \sf{m}},
L \in \mathbb{R}^{\sf{n} \times \sf{p}}$ be a feedback gain and 
an observer gain, respectively.
Let
$q_k \in \mathbb{R}^{\sf{p}} $
denote the quantized value of the sampled output $y_k$.
We denote by
$Q_1$ and $Q_2$ qunatization functions of the output estimate $\hat y$
and the control input $u$, respectively. 
For the plant $\Sigma_P$ in \eqref{eq:plant}, we construct
the following observer-based controller:
\begin{equation}
	\label{eq:observer_AQ}
	\Sigma_C': \begin{cases}
		\begin{aligned}
			\hat x_{k+1} &= A_d\hat x_k + B_d u_k + L(q_k - Q_1(\hat y_k))  \\
			\hat y_k &= C \hat x_k \\
			u_k &= -K \hat x_k
		\end{aligned}
	\end{cases}
\end{equation}
where $A_d$ and $B_d$ are defined as in \eqref{eq:discretization}.
We set the initial estimate $\hat x_0$ to be $\hat x_0 = 0$.
Compared with the controller $\Sigma_C$ in \eqref{eq:observer}, 
the controller $\Sigma_C'$ uses the quantized output estimate $Q_1(\hat y_k)$
instead of the original output estimate $\hat y_k$.
The control input $u(t)$ is produced as 
\begin{equation*}
	u(t) = Q_2(u_k)\qquad (kh \leq t < (k+1)h).
\end{equation*}
Note that the controller can
compute $\hat x_k$ and hence $\hat y_k, u_k$ 
at time $k-1$.
Fig.~\ref{fig:closed_loop_AQ} illustrates 
the closed-loop system we consider in this section.

\begin{rem}
	Instead of \eqref{eq:observer_AQ},
	we can use different controllers such as
	\begin{equation*}
		\Sigma_C'': \begin{cases}
			\begin{aligned}
				\hat x_{k+1} &= A_d\hat x_k + B_d Q_2(u_k) + L(q_k - Q_1(\hat y_k))  \\
				\hat y_k &= C \hat x_k \\
				u_k &= -K \hat x_k.
			\end{aligned}
		\end{cases}
	\end{equation*}
	The major reason to use \eqref{eq:observer_AQ} is that if $\hat x_0 = 0$, then
	we have
	\begin{equation}
		\label{eq:estimation_rep}
		\hat x_k = \sum_{\ell = 0}^{k-1} (A_d-B_dK)^{k-\ell} (q_{\ell} - Q_1(y_{\ell}))
	\end{equation}
	and hence $\hat y_k$ and $u_k$ can be described by $q_{\ell} - Q_1(y_{\ell})$,
	which makes encoding methods simple.
\end{rem}

\subsection{Output Encoding}
Suppose that we obtain an error bound $E_{k}$ such that 
$|y_k - Q_1(\hat y_k)| \leq E_{k}$.
A bound sequence $\{E_k\}$ satisfying this condition is obtained in 
Section 3.4.
Instead of \eqref{eq:quantization},
the encoder computes quantized measurements by 
dividing the hypercube 
\begin{equation}
	\label{eq:output_encoding_AQ}
	\left\{ y \in \mathbb{R}^{\sf p}:~| y  - Q_1(\hat y_k)|_{\infty} 
	\leq E_{k} \right\}
\end{equation}
into $N^{\sf p}$ equal boxes.
The difference between \eqref{eq:quantization} and \eqref{eq:output_encoding_AQ}
is the quantization center.
In \eqref{eq:quantization}, the encoder has the state estimate $\hat x_k$, and hence
the quantization center can be $\hat y_k$.
On the other hand, the encoder here employs  
the quantized output estimate $Q_1(\hat y_k)$ reported by the controller 
as the quantization center.
The rest of the output encoding is the same as in Subsection 3.2.

\subsection{Estimate and Input Encoding}
The controller sends the control input $u_k$ and the output estimate $\hat y_k$
to the plant side.
Suppose that we have bounds $E_{1,k}$ and $E_{2,k}$ such that 
$|\hat y_k| \leq E_{1,k}$ and $|u_k| \leq E_{2,k}$.
Such a bound sequence $\{(E_{1,k},E_{2,k})\}$ is obtained in 
Section 3.4.
The bounds and levels of 
the quantization of $\hat y, u$  
are given by $(E_{1,k}, N_1)$ and $(E_{2,k}, N_2)$. Namely,
the controller computes the quantized output estimate and the quantized input by 
dividing the hypercubes
\begin{align*}
	&\left\{ \hat y \in \mathbb{R}^{\sf p}:~| \hat y  |_{\infty} 
	\leq E_{1,k} \right\} \\         
	&\left\{ u \in \mathbb{R}^{\sf m}:~| u|_{\infty} 
	\leq E_{2,k} \right\}
\end{align*}
into $N_1^{\sf p}$ and $N_2^{\sf m}$ equal boxes and 
assigns a number in $\{1,\dots, N_1^{\sf p}\}$ and $\{1,\dots, N_2^{\sf m}\}$
to each divided box by a certain one-to-one mapping, respectively.
The decoder in the plant side 
generates $Q_1(\hat y_k)$  and $Q_2(u_k)$ 
equal to the center of the boxes
with number reported by the controller. 
Thus $Q_1(\hat y_k)$  and $Q_2(u_k)$ satisfy
\begin{align}
	\label{eq:quantization_error_hat_y_u}
	|\hat y_{k} - Q_1(\hat y_{k})| \leq \frac{E_{1,k}}{N_1},\quad
	|u_{k} - Q_2(u_{k})| \leq \frac{E_{2,k}}{N_2}.
\end{align}
Since $\hat x_0 = 0$, we can set the initial values
\[
E_{1,0} := 0,\quad E_{2,0} := 0.
\]

The encoding strategy \eqref{eq:output_encoding_AQ} 
of the output $y_k$ uses the quantized output estimate $Q_1(\hat y_k)$
as the quantization center, whereas
the quantization centers for
the output estimate $\hat y_k$ and the input $u_k$ are
the origin, which allows the plant side to have less computational resources.

\begin{rem}
	Although the encoder does not have to estimate output measurements,
	the bounds $E_k$, $E_{1,k}$, and $E_{2,k}$ should be computed 
	in the plant side.
	However, these bounds can be calculated
	by simple difference equations \eqref{eq:E_def_KAL_AQ}
	as shown in the next subsection.
	If components in the plant side do not have computational resources
	enough to implement those difference equations,
	the controller can send sufficiently accurate values of the bounds even
	in the presence of quantization,
	because the dimension of each bound is one.
\end{rem}

\subsection{Computation of Bound Sequence $\{(E_k,E_{1,k},E_{2,k})\}$}
The following theorem is an extension of 
Theorem \ref{thm:no_disturbances_non_deadbeat}.
\begin{thm}
	\label{thm:no_disturbances_non_deadbeat_AQ}
	Let Assumptions \ref{assum:initial_bound} and 
	\ref{assump:stabilizability_detectability} hold, and
	define $R$ and $\bar R$ as in \eqref{eq:R_def}.
	Let the observer gain $L$ and the feedback gain $K$
	satisfy
	$r(R) < 1$, $r(\bar R) < 1$ and 
	\begin{gather*}
		\|CR^{\ell}\| \leq M_0 \rho^{\ell},\quad 
		\|C\bar R^{\ell} L\| \leq M_1 \bar \rho^{\ell},\quad
		\|K\bar R^{\ell} L\| \leq M_2 \bar \rho^{\ell} \\
		\|CR^{\ell} 
		B_d \| \leq M_3 \rho^{\ell},\quad
		\|CR^{\ell} L\| \leq M_4 \rho^{\ell}
	\end{gather*}
	hold for some $M_0,M_1,M_2,M_3,M_4> 0$ and $\bar \rho \leq \rho < 1$.
	Define constants $\alpha_0$ and $\alpha_1$ by
	\begin{align}
		M &:= 
		(N-1) \cdot 
		\left(
		\frac{M_1M_4}{N_1} + \frac{M_2M_3}{N_2} 
		\right) \notag \\
		\beta_0 &:= 
		\rho + 
		\frac{N_1M_4 + (N-1)M_1}{NN_1} \notag \\
		\beta_1 &:=
		\frac{M}{N} \notag \\
		\alpha_0 &:= 
		\rho + \beta_0 \notag \\
		\alpha_1 &:=
		\beta_1-\rho \beta_0. \notag
	\end{align}
	If we pick $N,N_1,N_2 \geq 2$ so that 
	\begin{equation}
		\label{eq:F_AQ_general}
		F := 
		\begin{bmatrix}
			\alpha_0 & \alpha_1 \\
			1 &  0 
		\end{bmatrix}
	\end{equation}
	satisfies $r(F) < 1$, then
	the proposed encoding method with 
	a bound sequence $\{(E_k,E_{1,k}, E_{2,k})\}$ 
	defined by
	\begin{align}
		&E_{k+1} :=
		\begin{cases}
			M_0 E_{st} \rho + \left(M_4+\frac{(N-1)M_1}{N_1}\right)  \frac{E_0}{N}
			& k = 0 \\
			\beta_0 E_k + \beta_1 E_{k-1} & k =1\\
			\alpha_0 E_k + \alpha_1 E_{k-1} & k \geq 2
		\end{cases} \notag \\
		&E_{1,k+1} := \bar \rho E_{1,k} + \frac{(N-1)M_1}{N} E_k \hspace{67.5pt} k\geq 0 
		\notag  \\
		&E_{2,k+1} := \bar \rho E_{2,k} + \frac{(N-1)M_2}{N} E_k \hspace{67.5pt} k \geq 0.
		\label{eq:E_def_KAL_AQ}
	\end{align}
	achieves the exponential convergence of 
	the state $x$ and the estimate $\hat x$.
\end{thm}
\begin{pf}
	The error 
	$e_{k+1} = x_{k+1} - \hat x_{k+1}$ satisfies
	\begin{align}
		e_{k+1} &=
		R^{k+1} e_{0} + 
		\sum_{\ell = 0}^{k}
		R^{\ell} \big(
		L(y_{k-\ell} - q_{k-\ell}) \notag\\ 
		&\qquad \qquad \qquad - L(\hat y_{k-\ell} - Q_1(\hat y_{k-\ell}))\notag \\
		&\qquad \qquad \qquad - 
		B_d (u_{k-\ell} - Q_2(u_{k-\ell}))
		\big).
		\label{eq:e_all_quantize}
	\end{align}
	On the other hand,
	since $\hat x_0 = 0$, it follows that
	\begin{align*}
		\hat x_{k+1} =
		\sum_{\ell = 0}^k
		\bar R^{\ell} 
		L
		(q_{k-\ell} - Q_1(\hat y_{k-\ell})).
	\end{align*}
	Since $Q_1(\hat y_k)$ is the quantization center and
	$q_{k}$ is the quantization value,
	we have \[
	|q_{k} - Q_1(\hat y_{k})| \leq (N-1)\mu_k
	\] 
	for all $k \geq 1$. 
	Hence $\hat y_k$ and $u_k$ satisfy
	\begin{equation}
		\begin{array}{c}
			|\hat y_k| 
			\leq
			\frac{(N-1)M_1}{N}
			\sum_{\ell=0}^{k-1}
			\bar \rho^{\ell} E_{k-\ell -1} =: E_{1,k} \\[15pt]
			|u_k| 
			\leq
			\frac{(N-1)M_2}{N}
			\sum_{\ell=0}^{k-1}
			\bar \rho^{\ell} 
			E_{k-\ell -1} =: E_{2,k}  \label{eq:input_bound}
		\end{array}
	\end{equation}
	for all $k \geq 1$.
	Since $E_{i,k}$ in \eqref{eq:input_bound} satisfies
	\begin{equation*}
		E_{i,k+1} -\bar \rho E_{i,k} = 
		\frac{(N-1)M_i}{N} E_k, \qquad i =1,2, 
	\end{equation*}
	we obtain
	the difference equations \eqref{eq:E_def_KAL_AQ}
	for $E_{1,k}$ and $E_{2,k}$.
	
	On the other hand,
	for every $k \geq 0$,
	define $E_{k+1}$ by
	\begin{align}
		E_{k+1} :=
		M_0 E_{st} \rho^{k+1}  &+ 
		\sum_{\ell=0}^{k}
		\left(M_4+\frac{(N-1)M_1}{N_1}\right)\rho^{\ell}\mu_{k-\ell} \notag \\
		&+
		M
		\sum_{\ell=0}^{k-1}
		\sum_{i=0}^{k-\ell-1}
		\rho^{\ell+i} \mu_{k-\ell-i-1}. \label{eq:output_bound}
	\end{align}
	Combining \eqref{eq:quantization_error_hat_y_u},
	\eqref{eq:e_all_quantize}, and \eqref{eq:input_bound},
	we have that for all $k \geq 0$,
	\begin{align*}
		|y_{k+1} - Q_1(\hat{y}_{k+1})| 
		&\leq
		|y_{k+1} - \hat y_{k+1}| + 
		|\hat y_{k+1} - Q_1(\hat y_{k+1})| \\
		&\leq E_{k+1}.
		%
	\end{align*}
	Next we obtain the difference equation in \eqref{eq:E_def_KAL_AQ}
	from \eqref{eq:output_bound}.
	Since
	\begin{equation*}
		\sum_{\ell=0}^{k-1}
		\sum_{i=0}^{k-\ell-1}
		\rho^{\ell+i} \mu_{k-\ell-i-1} 
		=
		\sum_{\ell=0}^{k-1}
		\sum_{j=\ell}^{k-1}
		\rho^{j} \mu_{k-j-1},
	\end{equation*}
	it follows that 
	for every $k \geq 1$,
	$E_{k+1}$ in \eqref{eq:output_bound} satisfies
	\begin{align*}
		E_{k+1}
		&=
		\beta_0 E_k + 
		\beta_1 
		\sum_{\ell = 0}^{k-1}\rho^{\ell} E_{k-\ell-1}.
	\end{align*}
	We therefore have
	\begin{equation*}
		E_{k+1} - \rho E_k 
		=
		\beta_0 E_k + (\beta_1 - \beta_0\rho) E_{k-1}
	\end{equation*}
	for every $k \geq 2$.
	Thus, we obtain the
	difference equation \eqref{eq:E_def_KAL_AQ} for $E_k$.
	
	If we define a vector ${\bm \mu}_k$ by
	\begin{equation*}
		{\bm \mu}_k := 
		\begin{bmatrix}
			\mu_k \\ \mu_{k-1}
		\end{bmatrix}
	\end{equation*}
	and a matrix $F$ as in \eqref{eq:F_AQ_general},
	then we have the dynamics of ${\bm \mu}_k$, \eqref{eq:bmu_eq}.
	Hence $\mu_k$ exponentially decreases to zero 
	if and only if $F$ is Schur stable.
	Since the quantization errors of the input and the estimate 
	also exponentially decrease from \eqref{eq:E_def_KAL_AQ},
	the rest of the proof is the same as that of 
	Theorem \ref{thm:no_disturbances_non_deadbeat},
	and we therefore omit it.
	\hspace*{\fill} $\blacksquare$
\end{pf}

\begin{rem}
	Although we here generate $Q_1(\hat y_k)$ and $Q_2(u_k)$ separately,
	one can quantize $\hat y_k$ and $u_k$ simultaneously.
	This simultaneous quantization reduces 
	the computational cost of the controller, but
	the data-rate condition for stabilization becomes conservative.
	Therefore, we do not proceed along this line.
\end{rem}

\begin{rem}
	There always exist quantization levels $N$, $N_1$, and 
	$N_2$ such that
	the matrix $F$ in \eqref{eq:F_AQ_general} satisfies $r(F) < 1$.
	In fact,  as $N$, $N_1$, and $N_2$ increase to infinity, we have
	$\alpha_0 \to \rho$ and $\alpha_1 \to \rho^2$, and hence
	the eigenvalues of $F$ tend to $\rho < 1$.
\end{rem}

\section{Numerical Examples}
\subsection{Comparison of Data-Rate Conditions}
First we consider the quantization of only the plant output and
compare the data-rate conditions of three types of observers:
the steady-state Kalman filter \eqref{eq:CRL_bound} 
with process noise covariance $10^{-3}$ and
measurement noise covariance $10^{-5}\text{diag}(1,1)$, 
the deadbeat observer \eqref{eq:F_spectrum}, and the pseudo-inverse observer  \eqref{eq:output_liberzon}.
In addition to the output feedback case, we also investigate 
the state encoding case \eqref{eq:state_liberzon}.
In Table \ref{table:comparison}, we show the comparison of the minimum
quantization level for exponential convergence, which are
$N^{\sf p}$ in the output feedback case and
$N^{\sf n}$ in the state feedback case.
Note that steady-state Kalman filter and the deadbeat observers
are represented by a linear time-invariant state equation
but pseudo-inverse observers does not.

The first example is an inverted pendulum whose dynamics is given
by \eqref{eq:plant} with
\begin{gather}
	A := 
	\begin{bmatrix}
		0 & 1 & 0  & 0 \\
		0  &-20.06 &  53.26 &  -1.096 \\
		0  & 0  &  0  &  1\\
		0  & -20.01 & 98.41 &   -2.025
	\end{bmatrix},\quad
	B :=
	\begin{bmatrix}
		0 \\  35.28 \\ 0 \\ 35.18
	\end{bmatrix} \label{eq:inv_pen} \\
	C :=
	\begin{bmatrix}
		1 &  0 & 1 & 0
	\end{bmatrix}. \notag 
\end{gather}
The state $[x_1~~x_2~~x_3~~x_4] =: x$
are the arm angle, the arm angular velocity, 
the pendulum angle, and the pendulum angular velocity.
The input $u$ is the motor voltage.
Additionally, we borrow a 2-mass motor drive with one output and three states
from \cite{Ji1995},
a pneumatic cylinder with one output and three states from \cite{Kimura1996}, and
a batch reactor with two output and four states from \cite{Rosenbrock1972}.

In Table \ref{table:comparison},
we see that the Kalman filter requires less data-rate than the deadbeat observer 
and the pseudo-inverse observer
for
the 2-mass motor drive and the pneumatic cylinder.
This is because the motor drive and the cylinder have their unstable poles
only on the imaginary axis. Hence the observer gain of the Kalman filter is small,
which decreases $M$ in \eqref{eq:CRL_bound}.
Although deadbeat observers and
pseudo-inverse observers have the same property:
finite-time state reconstruction in the
idealized situation without quantization,
the data-rate condition \eqref{eq:output_liberzon} by pseudo-inverse observers 
is better than that \eqref{eq:F_spectrum} by deadbeat observers.
This is because pseudo-inverse observers employ output measurements directly
for state reconstruction, whereas deadbeat observers summarize 
output information by their states. Moreover,
compared with the state feedback case \eqref{eq:state_liberzon},
the output feedback case requires small data sizes in most numerical examples
because the state dimension ${\sf n}$ and 
the output dimension ${\sf p}$ satisfy ${\sf n} > {\sf p}$.

\begin{table*}[!hbt]
	\caption{Minimum quantization levels for exponential convergence.}
	\label{table:comparison}
	\centering
	\begin{tabular}{ccccc} \toprule
		& Kalman filter \eqref{eq:CRL_bound}
		& Deadbeat observer \eqref{eq:F_spectrum}
		& Pseudo-inverse observer \eqref{eq:output_liberzon}
		& State encoding \eqref{eq:state_liberzon}\\ \hline \hline	   
		Inverted pendulum ($h=0.03$) & 
		$5^2=25$ & $4^2=16$  &$4^2=16$ & $4^4 = 256$ \\
		2-mass motor drive ($h=10^{-3}$) & $5^1 = 5$  & $7^1 = 7$ & $3^1 = 3$ & $3^3 =27$  \\
		Pneumatic cylinder ($h=10^{-3}$)& $2^1 = 2$ & $7^1 = 7$ &$3^1=3$ & $2^3 = 8$ \\
		Batch reactor ($h=0.1$) &  $23^2 = 529$ & $4^2 = 16$   &$4^2 = 16 $ & $3^4 = 81$ \\   \bottomrule 
	\end{tabular}
\end{table*}

\subsection{Time response of Inverted Pendulum}
Consider again the inverted pendulum in the previous subsection.
Next we compute the time response of the inverted pendulum described by \eqref{eq:inv_pen}
with considering quantization of the plant input and output.
The controller is assumed to send to the encoder the quantized value
of the output estimate for quantization centers.
Let the sampling period be $h = 0.03$ sec. We set 
the feedback gain $K$ to be
the quadratic regulator whose
weighting matrices of the state and the input are
$\text{diag}(100,0,300,0)$ and
$1$, respectively. 
The observer gain $L$ is
the Kalman filter whose 
covariances of the process noise and measurement noise are
$10^{-3}$ and $10^{-5}\text{diag}(1,1)$, respectively.

From Theorem \ref{thm:no_disturbances_non_deadbeat_AQ},
the state $x$ exponentially converges to the origin
under
the proposed encoding strategy with $(N,N_1,N_2) = (151,301,1601)$ for which
$F$ in \eqref{eq:F_AQ_general} satisfies $r(F) = 0.8845$.

\begin{figure}[b]
	\centering
	\includegraphics[width = 8cm,clip]{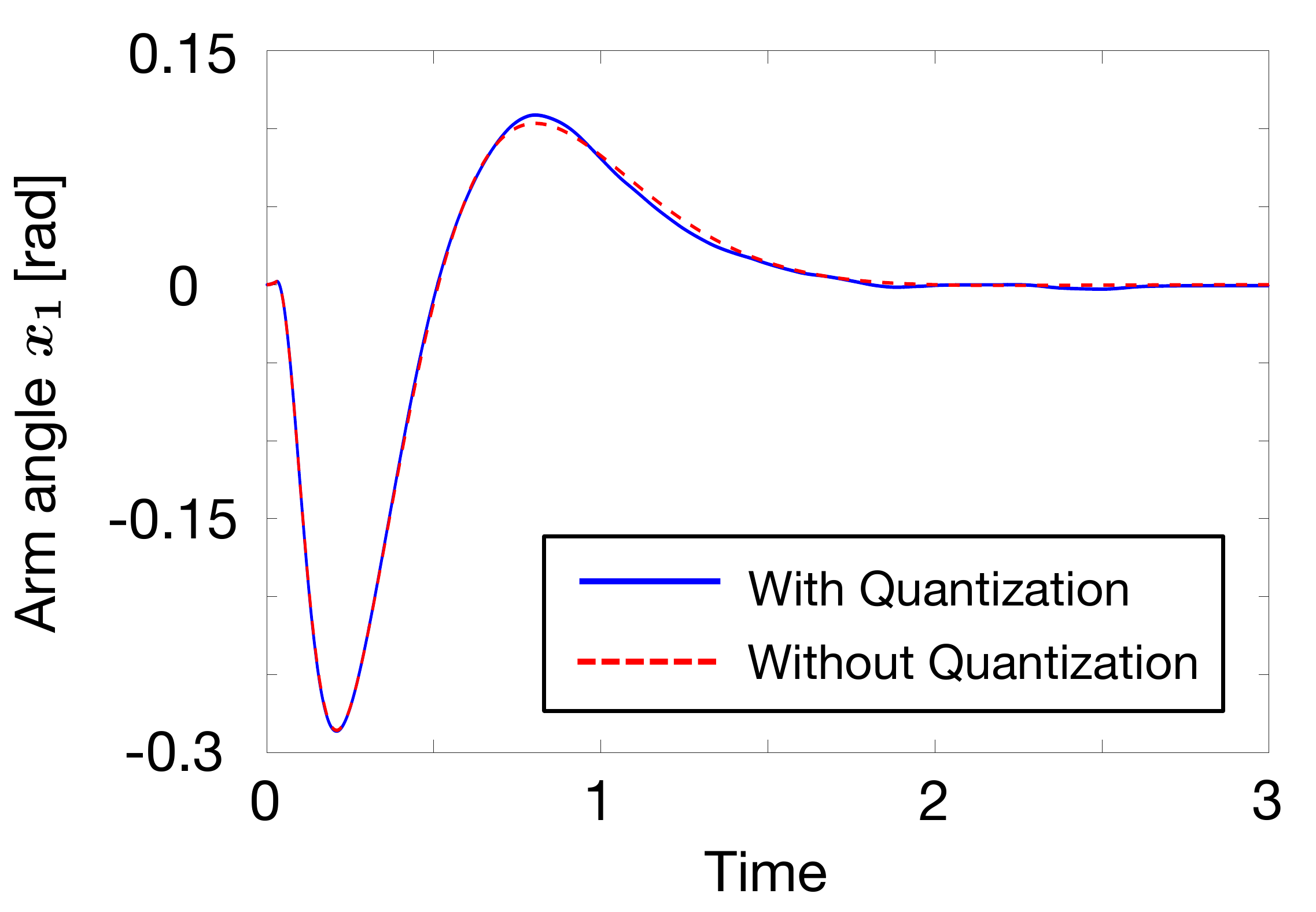}
	\caption{Arm angle $x_1$ with/without quantization. }
	\label{fig:arm_angle}
\end{figure}

\begin{figure}[b]
	\centering
	\includegraphics[width = 8cm,clip]{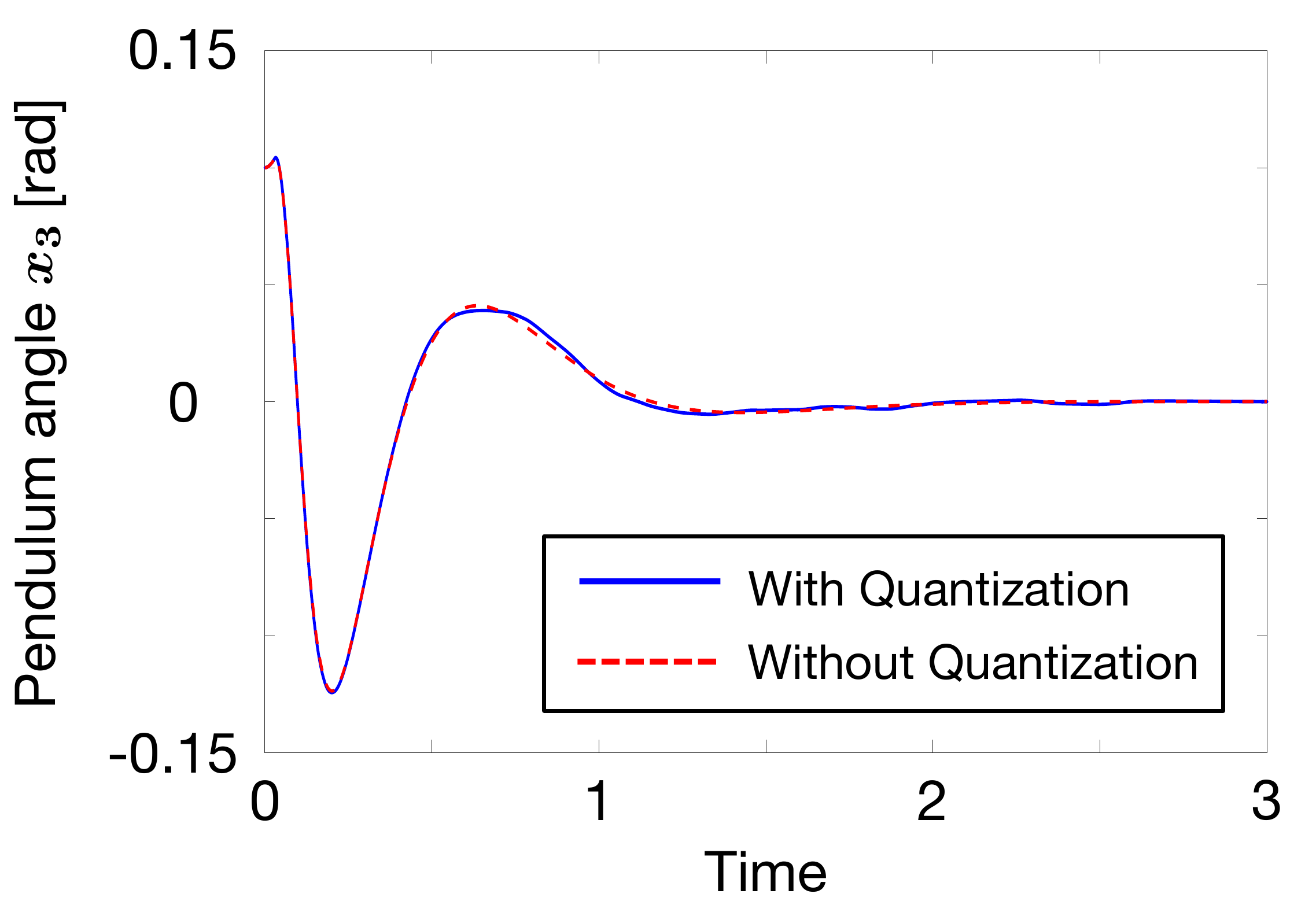}
	\caption{Pendulum angle $x_3$ with/without quantization. }
	\label{fig:pen_angle}
\end{figure}

\begin{figure}[b]
	\centering
	\includegraphics[width = 8cm,clip]{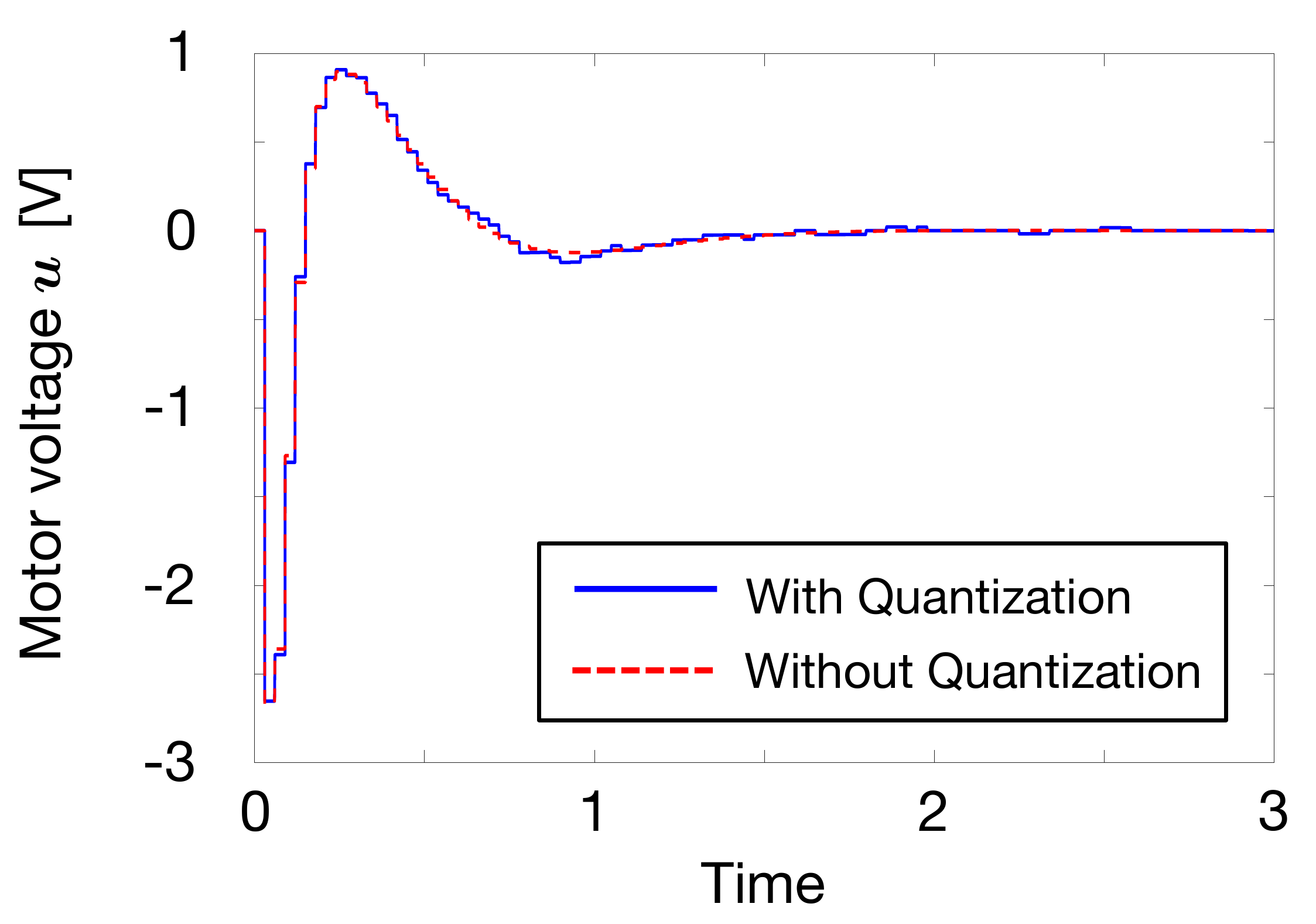}
	\caption{Motor voltage $u$ with/without quantization. }
	\label{fig:pen_input}
\end{figure}

Supposing that we obtain an initial state bound $E_0 = 0.15$,
we compute
a time response for the initial state
$x(0) = [0~~0~~0.1~~0]^{\top}$.
The plot of the arm angle $x_1$ and the pendulum angle $x_3$ 
is  in Figs.~\ref{fig:arm_angle} and \ref{fig:pen_angle}.
Fig.~\ref{fig:pen_input} illustrates the motor voltage $u$.
From Figs.~\ref{fig:arm_angle} and \ref{fig:pen_angle},
we observe that the arm and pendulum angles decrease to zero
in the presence of
three types of quantization errors.
Since the initial estimate $\hat{x}_0 = 0$, 
the quantization errors of the output estimate and the input are small 
at first.
In fact, the output estimate bound $\{E_{1,k}\}$ and 
the input bound $\{E_{2,k}\}$ take the maximum value at about time $t = 0.45$, and
the effect of 
the quantization errors appears from time $t = 0.5$ in Figs.~\ref{fig:arm_angle},
\ref{fig:pen_angle}, and \ref{fig:pen_input}.

\section{Conclusion}
We studied quantized output feedback stabilization by
Luenberger observers.
Data-rate conditions
for general Luenberger observers are characterized by
the spectral radius of the system matrix of the error dynamics.
On the other hand, a data-rate condition for deadbeat observers 
is determined by the behavior of the error dynamics for $\eta$ steps, where
$\eta$ is the observability index of the plant. 
The proposed encoding method was also extended to case where
both of the plant input and output are quantized and where
the encoder does not have an estimator for generating quantization centers.
Future work involves addressing more general systems such as
nonlinear systems and switched systems.

\end{document}